\documentclass[12pt,a4paper]{article}

\usepackage[T1]{fontenc}
\usepackage[utf8]{inputenc}
\usepackage{lmodern}
\usepackage{amsmath,amssymb}
\usepackage{graphicx}
\usepackage{booktabs}
\usepackage{array}
\usepackage{multirow}
\usepackage{caption}
\usepackage{geometry}
\usepackage{hyperref}
\usepackage{natbib}
\usepackage{setspace}
\usepackage{titlesec}
\usepackage{microtype}
\usepackage{rotating}

\title{\textbf{Density-functional theory calculation of hydrogen solubility in cubic silicon carbide at finite temperatures}}

\author{
  Jonathan S.\ Evarts\textsuperscript{a,*} \and
  Anne Chaka\textsuperscript{a} \and
  Towfiq Ahmed\textsuperscript{a}
}

\date{
  \small\textsuperscript{a}Pacific Northwest National Laboratory, Richland, WA 99354, USA\\[0.5em]
}

\begin{document}

\maketitle

\footnotetext[1]{Corresponding author: jonathan.evarts@pnnl.gov, +1\,(509)\,372-7873}

\begin{abstract}
An \textit{ab initio} framework using density-functional theory has been developed to predict hydrogen
solubility in both pristine and defective $\beta$-SiC. This study is motivated by the critical need
for accurate hydrogen permeation models in fusion reactor designs, where predicting hydrogen
permeation, which is inherently dependent on its solubility, through tritium permeation barrier
(TPB) materials is essential. Although silicon carbide is one of the leading candidates for TPBs,
experimental permeation values vary widely due to differences between ideal single crystals and
real, defect-containing materials, as well as variations in measurement techniques. Here, first
principles calculations are employed to quantify the effects of interstitials, vacancies, and
nonstoichiometric (amorphous) structures on hydrogen behavior in $\beta$-SiC, thereby clarifying
the discrepancies between ideal and defective systems. Our results show that hydrogen solubility
is significantly enhanced in carbon-rich nonstoichiometric amorphous structures and silicon
vacancies compared to hydrogen occupying interstitial sites in pure $\beta$-SiC.
\end{abstract}

\section{Introduction}
\label{sec:intro}

Effective tritium permeation barriers (TPB) are crucial for advancements in fusion energy
technology and have been identified as a critical problem for first-wall materials \cite{Hassan2021,
Ivanov2024, Luscher2013, Houben2020, Linsmeier2017, Chikada2011}. Understanding tritium behavior in
first-wall materials is important for several reasons: (1)~preventing tritium release to the
environment, (2)~accurately predicting fuel loading in the plasma, and (3)~preventing embrittlement
of support structures \cite{Linsmeier2017, Oya2006}. Among the promising TPB first-wall materials
is SiC \cite{Sizyuk2024}, in which studies have investigated several phases of SiC synthesized in
a variety of ways, including single-crystal SiC \cite{Causey1978}, chemical vapor deposition (CVD)
$\beta$-SiC \cite{Causey1978, Causey1995}, sintered KT-SiC\footnote{KT-SiC is a self-bonded SiC
formerly produced by the Carborundum Company.} \cite{Verghese1979}, and CVD-SiC coated on metals
\cite{Chikada2011, Sinharoy1984}. Despite extensive testing, experimental permeation measurements
have yielded inconsistent results that vary by up to six orders of magnitude, with CVD $\beta$-SiC
and single crystal $\alpha$-SiC exhibiting the lowest permeation
\cite{Chikada2011, Causey1978, Causey1995, Verghese1979, Sinharoy1984, Causey1993, Causey2012,
Tam1995, Minami2007, Yamamoto2016, Wright2015}. Variations in permeability are attributed to
experimental methods, and sample preparation differences, including factors such as CVD parameters,
substrate materials, and irradiation-induced defects \cite{Katoh2006, Price1973, Senor2003,
Sprouster2021, Snead2012}.

Irradiation, for instance, produces high concentrations of Frenkel loops, dislocation networks,
and swelling at $<2$\,dpa, which may increase H trapping sites in the material
\cite{Katoh2006, Price1973, Senor2003, Sprouster2021, Snead2012, Sun2017, Yano1988}. Trapping
occurs at defects where the binding energy is sufficiently strong to withstand high temperature.
Stronger H binding in trap sites means that thermal conditions must be met to overcome these
barriers, thereby affecting H mobility. Trapping can appear to increase solubility, as more H
atoms temporarily reside at these defect sites. However, solubility fundamentally represents the
equilibrium concentration of H atoms uniformly dissolved across the SiC lattice, governed by the
activation energy for incorporation into the lattice.

While H permeation has been extensively measured, with reported values differing by several orders
of magnitude, H solubility in SiC remains less well-explored. Early studies, such as those by
Causey et al.\ \cite{Causey1978}, faced challenges in reaching equilibrium saturation due to sample
thickness, complicating the accurate solubility calculations from diffusivity measurements.
Follow-on studies measured solubility at
$1.36\times10^{-2}\,\exp(58.8\,\text{kJ\,mol}^{-1}/RT)$\,mols\,H$_2$\,m$^{-3}$\,MPa$^{-1/2}$
\cite{Causey1993}. H concentration measurements at 0.1\,MPa performed on SiC$_\text{f}$/SiC
composites found
$1.9\times10^{3}\,\exp(-30.3\,\text{kJ\,mol}^{-1}/RT)$\,mols\,m$^{-3}$ \cite{Esteban2002}.

As previously noted, trap sites significantly influence solubility measurements due to their
tendency to capture H atoms. Density-Functional Theory (DFT) calculations have shown that
interstitials and vacancies in SiC act as trap sites, with certain configurations like the silicon
($V_{\text{Si}}$) and carbon vacancies ($V_{\text{C}}$) forming particularly stable H complexes
\cite{Aradi2001, Bernardini2004, Iwamoto2015, Kaukonen2003, Torpo2001, Wang2020, Zywietz1999}.
However, there is a notable gap in existing research regarding how binding energies vary as a
function of temperature or hydrogen partial pressure. Addressing this deficiency is crucial to
developing comprehensive models of tritium behavior in first-wall materials. This distinction
underscores the necessity to understand both solubility and trapping mechanisms, each characterized
by different energy profiles as a function of temperature and hydrogen partial pressure, to
accurately predict H permeation.

Given the discrepancies observed in permeability results and the lack of solubility data for SiC,
this study aims to develop a model predicting solubility while considering microstructural
influences. Our focus is on a first principles assessment of interstitials and defects impact on H
solubility to resolve experimental inconsistencies. Additionally, we evaluate defect trapping
capacities across varying temperatures and pressures, given that stronger traps associated with
deeper potential wells can significantly reduce permeability. By implementing DFT calculations
complemented with JANAF thermochemical data (\textit{ab initio} thermodynamics), this study will
begin to address the lack of research into binding energies under varying conditions, as well as
calculating H solubility as a function of temperature, H partial pressure, and defect densities
in $\beta$-SiC, including within amorphous regions, areas not yet extensively explored
\cite{Causey1993, Nadaoka2009}.

\section{Computational Methods}
\label{sec:methods}

Structures and total energies were calculated using DFT as implemented in DMol$^3$
\cite{Delley2000} using the generalized gradient approximation (GGA) of Perdew-Burke-Ernzerhof
(PBE) to the density functional. Calculations use a self-consistent field convergence of
$10^{-5}$\,Hartrees, and a real-space cutoff for the numerical, atom-centered basis functions of
5.1\,\AA. A $2\times2\times2$ supercell with 64 atoms, and a $4\times4\times4$ $k$-point mesh was
found to be converged with respect to minimizing interactions between H atoms and point defects,
and for phonon frequencies. Geometry optimization calculations were carried out using the
Broyden-Fletcher-Goldfarb-Shanno algorithm by minimizing the Hessian to an energy convergence of
$10^{-5}$\,Hartrees. Supercell model geometries were optimized and converged to within
0.01\,eV.

\begin{figure}[htbp]
  \centering
  \includegraphics[width=\linewidth]{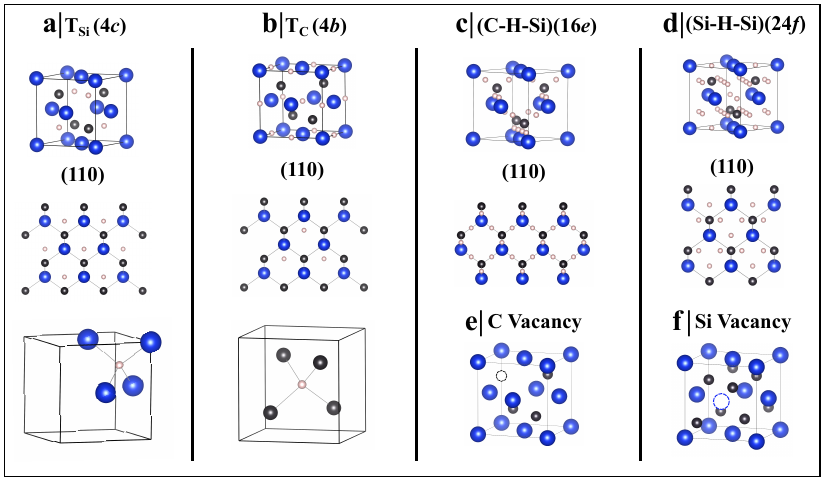}
  \caption{Visualization of crystal structures and interstitials in $\beta$-SiC. Silicon atoms
    are blue, carbon atoms are black, and H is pink. Wyckoff interstitial sites (a)~$4c$,
    (b)~$4b$, (c)~$8c$, and (d)~$24f$. (e,f)~are the carbon vacancy and silicon vacancy,
    respectively.}
  \label{fig:crystal}
\end{figure}

To calculate the solubility of H in $\beta$-SiC, several types of H-binding sites were generated
in both the perfect and defective crystal structures. $\beta$-SiC is a zinc blende structure
($F\overline{4}3m$) with Si atoms residing in the Wyckoff $4a$ position $(0,0,0)$ and the C atoms
occupying $\frac{1}{2}$ of the Wyckoff $4c$ positions
$(\frac{1}{4},\frac{1}{4},\frac{1}{4})$. Each C is tetrahedrally coordinated by four Si atoms,
and vice versa. There are four types of unique sites for H interstitial binding: the unoccupied
$4c$ position (T$_{\text{Si}}$), which is coordinated by four Si atoms; the $4b$ position
$(\frac{1}{2},\frac{1}{2},\frac{1}{2})$ (T$_{\text{C}}$) coordinated by four C atoms; the $16e$
position $(\frac{1}{8},\frac{1}{8},\frac{1}{8})$, $(\frac{1}{8},\frac{5}{8},\frac{5}{8})$
located between a Si and a C; and the $24f$ position $(\frac{1}{4},\frac{1}{4},0)$ lying along
the $\langle100\rangle$ direction, as shown in Figure~\ref{fig:crystal}. In the cubic SiC
structure, all Si and C atoms are equivalent, so only one type of vacancy site was considered for
each. For the DFT binding energy calculations, one H atom was placed in either one of four
interstitial sites, a $V_{\text{Si}}$ or a $V_{\text{C}}$ within the supercell. For single
$V_{\text{Si}}$ and $V_{\text{C}}$ sites in a $2\times2\times2$ supercell, the vacancy
concentrations are $4.7\times10^{20}$\,cm$^{-3}$, which is approximately equivalent to the point
defect concentration of $2.41\times10^{18}$\,cm$^{-3}$ measured using positron annihilation
\cite{Brauer1996}.

\begin{figure}[htbp]
  \centering
  \includegraphics[width=\linewidth]{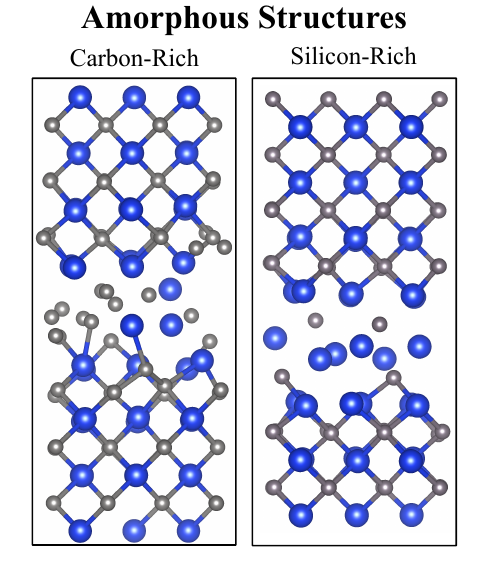}
  \caption{Amorphous structures built for carbon-rich and silicon-rich structures. The amorphous
    structures occupy approximately 25\,vol.\% of the structure.}
  \label{fig:amorphous}
\end{figure}

In SiC, highly disordered grain boundaries or amorphous structures can be generated as a result
of irradiation damage or during low-temperature CVD processes. To model examples of these regions,
structures were generated by removing several atoms (Si atoms for C-rich and C atoms for Si-rich)
in a band along the $(001)$ plane and optimizing the structure (Figure~\ref{fig:amorphous}). The
resultant amorphous structure occupied $\approx25$\,vol.\% of the supercell. This fraction was
chosen to minimize computational time while generating a reasonable amorphous content
concentration. Previous works for H solubility in metals have shown that when interactions between
H atoms become non-negligible, solubility deviates from experimental values \cite{Lee2015}. To
minimize H interactions in this work, only one H atom was placed in the amorphous region of the
supercell at a time. Due to the amorphous nature of the generated layer, the atomic configuration
exhibits significant randomness, resulting in a large number of possible H positions. A subset of
these positions were selected based on a qualitative assessment of the atomic structure to
represent a range of solubilities that may result from an amorphous structure. It should be noted
that this is a limited sampling of potential amorphous structures and not intended to be an
exhaustive or statistical examination of all possibilities. Our intent is to provide an initial
quantitative estimate of the potential impact of amorphous regions in SiC relative to other types
of defects and the perfect crystal.

H solubility is determined through a modification of the approach used in Ref.~\cite{Lee2015}
utilizing DFT total energy and phonon calculations. An important distinction in this study
compared to the method outlined in Ref.~\cite{Lee2015} is the removal of the electronic
degeneracy summation term, which is less relevant for semiconductors compared to metals at high
temperatures. Modifications from Ref.~\cite{Lee2015} also include substituting the free energy of
formation ($\Delta G_f$), as calculated using \textit{ab initio} thermodynamics described below,
in the partition function Boltzmann factor (see Eqs.~(S1) and (S2) in the Supplemental
Information, SI). Pressures from 0.02\,Pa ($1\times10^{-4}$\,mbar) to 0.1\,MPa (1\,bar) are used
to quantify partial pressure effects from 0 to 1000\,K. This work assumes that atomic H dissolves
in the solid at static partial pressure.

To validate the calculation methods described below, enthalpy of formations for crystalline
$\beta$-SiC were calculated and compared to NIST-JANAF thermochemical tables \cite{Chase1998}.
DFT calculated values were $\Delta H_{0\,\text{K}} = -72.27$\,kJ\,mol$^{-1}$, which are in good
agreement with NIST-JANAF values $\Delta H_{0\,\text{K}} = -72.22$\,kJ\,mol$^{-1}$.

\subsection{\textit{Ab Initio} Thermodynamics}
\label{sec:abinitio_thermo}

Calculating properties such as hydrogen solubility requires incorporating the effects of the
environment in equilibrium with a material, such as temperature and partial pressure of H$_2$,
$p_{\text{H}_2}$. To incorporate the effect of the external environment at finite temperatures
and pressures we employ the method of \textit{ab initio} thermodynamics \cite{Chaka2014,
Reuter2001, Wang2000, Zhang2004}. For a solid matrix, the Gibbs free energy $G = H - TS$, where
$H = U + pV$, is related to the DFT calculated total energies ($E_{\text{total}}$) through the
Helmholtz free energy ($F$), where $F = U - TS$ \cite{Chaka2014, Reuter2001, Wang2000, Mason2010}.
Combining and rewriting, the following Gibbs free energy expression is found:
\begin{equation}
  G(T,p) = U - TS(T) + pV ,
  \label{eq:gibbs}
\end{equation}
where $p$ is the pressure, $V$ the volume, $T$ the temperature, and $S$ the entropy. $U$ is the
internal energy and is equal to $E_{\text{total}}$ plus the vibrational contribution ($E_{\text{vib}}$).
At 0\,K the vibrational contribution is equal to the zero-point energy; at finite temperature the
vibrational contribution to the free energy is calculated from first-principles phonon dispersion.
Thus, $E_{\text{vib}}$ is the zero-point energy (ZPE) at 0\,K plus the vibrational entropy and
enthalpy at finite temperature and pressure. For convenience using DFT calculations, we choose
0\,K as the reference thermodynamic state. Considering that DFT calculations are performed at
constant volume and SiC is an incompressible material, the $pV$ term is ignored and the expression
for $G$ becomes:
\begin{equation}
  G(T,p) = E_{\text{total}}^{0\,\text{K},p} + E_{\text{vib}} .
  \label{eq:gibbs2}
\end{equation}

To calculate the free energy of formation $\Delta G_f$, the chemical potential of the component
atoms are related to $G$ through the relation $G = \sum_i N_i \mu_i$, where $\mu_i$ is the
chemical potential of constituent $i$ and $N_i$ is the number of atoms. $\mu_i$ is obtained from
the sum of the 0\,K reference state chemical potential $\mu_i^0$ (total energy and ZPE
calculations) plus the change in chemical potential from 0\,K to a given temperature and pressure
$\Delta\mu_i(T,p)$:
\begin{equation}
  \mu_i(T,p) = \mu_i^0 + \Delta\mu_i(T,p) .
  \label{eq:chempot}
\end{equation}
Then $\Delta G_f$ can be calculated as:
\begin{equation}
  \Delta G_f(T,p) = G(T,p) - \sum_i N_i \mu_i^0 + \sum_i N_i \Delta\mu_i(T,p) .
  \label{eq:dGf}
\end{equation}

At equilibrium the chemical potential of a species in all phases must be equal. Relating the H
chemical potential in the gas phase ($\Delta\mu_{\text{H}}$) to a particular pressure can be
determined by treating the surrounding gas phase environment as an ideal gas \cite{Reuter2001},
Eq.~(\ref{eq:muH}). Enthalpy and entropy values tabulated in the JANAF thermochemical data tables
\cite{Chase1998} are used to calculate $\mu_\text{H}$ for desired pressures using the ideal gas
law, where $p^o$ is the pressure of the gas species, $k$ is Boltzmann's constant, and $T$ is
temperature \cite{Wang2000}:
\begin{equation}
  \mu_{\text{H}}(T,p) = \mu_{\text{H}}(T,p^o) + \frac{1}{2} k T \ln\!\left(\frac{p}{p^o}\right) .
  \label{eq:muH}
\end{equation}

\subsection{Solubility}
\label{sec:solubility_method}

Solubility is calculated using a modified method from Lee et al.\ \cite{Lee2015}. In this model,
solubility is derived by equating the chemical potentials of gaseous H$_2$ and dissolved atomic
hydrogen, each expressed via its partition function $q_{\text{sol}}$ and $q_{\text{gas}}$
(Eqs.~(S1) and (S2) in SI), respectively. Configurational entropy is incorporated through a
combinatorial term accounting for hydrogen distribution among interstitial sites, simplified using
Stirling's approximation, yielding an expression for solubility and recovering Sieverts' law in
the dilute limit:
\begin{equation}
  \theta = \frac{q_{\text{sol}}}{\dfrac{q_{\text{gas}}\, PV}{k_B T} + q_{\text{sol}}} ,
  \label{eq:solubility}
\end{equation}
where $P$ is the H partial pressure, $V$ is the volume of H gas, $k_B$ is the Boltzmann constant,
and $T$ is the temperature in K, while $\theta$ is expressed as the ratio of H atoms to available
sites ($N_{\text{sites}}/N_{\text{atoms}}$); the appropriate conversion factor is applied (see
Table~S2 in SI). The amorphous structures built in this work do not contain a fixed number of
sites due to their random atomic arrangements. Considering that the amorphous structure has no
symmetry or equivalent positions, a conversion factor of 1 site per $N_{\text{atoms}}$ in the
supercell was chosen. As previously discussed, the amorphous volume is 25\,vol.\% of the supercell
structure; thus, solubility results are strictly defined as H solubility in a material with an
amorphous content of 25\,vol.\%. The same holds true for point defects. In this case, H
solubility results are calculated for a material with $4.7\times10^{20}$ defects\,cm$^{-3}$. In
this work, we extend the approach by Lee et al.\ \cite{Lee2015} by incorporating the free energy
of the system at finite temperatures—calculated through \textit{ab initio} thermodynamics—where
the Gibbs free energy $\Delta G_f$ (Eq.~\ref{eq:dGf}) is incorporated into the Boltzmann factor
term within the partition functions (Eqs.~(S1) and (S2) in SI).

\section{Results}
\label{sec:results}

Result subsections are divided into binding energies, free energies of formation, and solubility
calculations. Results are based upon H being placed within an interstitial site, a vacancy, or an
amorphous region. For H interstitial binding energy calculations, the reference state is the
crystalline structure. For H placed in defects, such as a vacancy or amorphous structure, the
reference state is the defective crystal, e.g.\ a crystal lattice containing a vacancy. Results
are indicative of a H atom, under specific pressures and temperatures, integrated into the SiC
structure. Negative values for binding energy and free energy of formation calculations indicate
trapping due to the favorable exothermic reaction.

\subsection{Hydrogen Binding Energies in SiC}
\label{sec:binding}

In $\beta$-SiC there are four distinct types of interstitial sites, as previously discussed and
shown in Figure~\ref{fig:crystal}. Results show the most favorable interstitial site is the
T$_{\text{Si}}$ site at an endothermic binding energy of 2.87\,eV. The T$_{\text{C}}$ site has a
binding energy of 3.22\,eV. H binding energies for bond center locations Si--H--Si and C--H--Si
are 2.93\,eV and 3.14\,eV; however, if a vacancy is imposed adjacent to the interstitial site,
the energy is lowered by up to 3.10\,eV due to the additional unpaired electron. A H in a
C--H--Si bond configuration results in H offsetting slightly to be closer to an adjacent C atom
with a distance of 1.12\,\AA. In the Si--H--Si bond, the Si--H bond lengths are 1.724\,\AA\ and
1.722\,\AA, forming a $179.64^\circ$ bond angle.

Vacancy formation in SiC exhibits distinct trends (see Table~\ref{tab:defects}), with $V_{\text{C}}$
being more readily formed than $V_{\text{Si}}$. These results show that the $V_{\text{Si}}$ has a
formation energy of 7.89\,eV and remains stable against decomposition into a $V_{\text{C}}$
(4.78\,eV) plus a carbon antisite ($C_{\text{Si}}$) (4.26\,eV), in contrast to 4H-SiC where the
$V_{\text{Si}}$ is metastable and tends to decompose into $V_{\text{C}} + C_{\text{Si}}$
\cite{Aradi2001}. This difference in formation energies suggests that $V_{\text{C}}$ are more
likely to form, potentially leading to C segregation at grain boundaries and Si-enriched grain
interiors. Higher formation energy defects, such as the $V_{\text{Si}}$, are less likely to
exist, but if they do exist and have strong H binding energies, they can significantly impact
solubility depending on their concentration.

With respect to hydrogen trapping, our calculations show that the $V_{\text{Si}} + \text{H}$
complex is favored over the $V_{\text{C}} + \text{H}$ complex by 0.91\,eV. Moreover, the H
binding energy in a $V_{\text{Si}}$ defect varies strongly with the local bonding environment.
When H bonds directly with a C within the $V_{\text{Si}}$, the binding energy ranges from
$-1.88$ to $-1.71$\,eV, while in configurations where bond formation is absent due to excessive
H--neighbor distances and localized orbitals, the binding energy is reduced to $-1.00$ to
$-0.97$\,eV. In the case of $V_{\text{C}} + \text{H}$, H is observed to migrate 0.60\,\AA\
(along the specified crystallographic direction) to form a Si--H--Si bond with a comparatively
low binding energy of $-0.06$\,eV. The stronger H binding energy in a $V_{\text{Si}}$ compared
to a $V_{\text{C}}$ comes from the higher formation energy of the $V_{\text{Si}}$. Since forming
a $V_{\text{Si}}$ requires a much larger energetic penalty, its creation leads to a deeper
potential energy well and thus a much stronger (more negative) binding energy. This pronounced
contrast in trapping behavior between vacancy (more exothermic) and interstitial sites emphasizes
the complex role that defect structure plays in governing hydrogen behavior in SiC.

\begin{sidewaystable}[htbp]
\centering
\caption{Comparison of defect formation energies in $\beta$-SiC at 0\,K, referenced to a
  defect-free crystal and to a defective crystal. Energies calculated in this work are compared
  to Aradi et al.\ \cite{Aradi2001} (4H-SiC), Zywietz et al.\ \cite{Zywietz1999} (3C-SiC), and
  Bernardini et al.\ \cite{Bernardini2004} (3C-SiC). The 32-atom supercell from Aradi et al.\
  includes zero-point energies. $\mu_\text{H}$ is the chemical potential of H; reasonable values
  are $-13$ to $-17$\,eV.}
\label{tab:defects}
\small
\begin{tabular}{lcccccc}
\toprule
 & \multicolumn{4}{c}{Formation Energy (eV) at 0\,K [Crystal]} & \multicolumn{1}{c}{Formation Energy (eV) at 0\,K [Defect]} \\
\cmidrule(lr){2-5}\cmidrule(lr){6-6}
Defect & This work ($\beta$-SiC) & Aradi \cite{Aradi2001} & Zywietz \cite{Zywietz1999} & Bernardini \cite{Bernardini2004} & This work (64-atom) \\
\midrule
\multicolumn{6}{l}{\textit{Interstitials}} \\
H$_{\text{T}_{\text{Si}}}$ & 2.87 & -- & -- & -- & -- \\
H$_{\text{T}_{\text{C}}}$ & 3.22 & -- & -- & -- & -- \\
Si--H--Si & 2.93 & -- & -- & -- & 0.02 (2.93)$^e$ \\
Si--H--C  & 3.14 & -- & -- & -- & -- \\
\midrule
\multicolumn{6}{l}{\textit{Vacancies}} \\
$V_{\text{C}}$ & 4.78 & 4.49 & 4.30 & 3.84$^a$, 4.50$^b$ & -- \\
$V_{\text{Si}}$ & 7.89 & 8.16 & 8.69 (8.45)$^c$ & 8.78$^a$, 8.12$^b$ & -- \\
$V_{\text{C}}+\text{H}$ & -- & $-10.76 - \mu_\text{H}$ & -- & -- & $-0.06$ $(0.90)^d$ \\
$V_{\text{Si}}+\text{H}$ & -- & $-9.36 - \mu_\text{H}$ & -- & -- & $-0.97$ \\
\midrule
\multicolumn{6}{l}{\textit{Amorphous Structures}} \\
H$_{\text{GB}}$ (C-rich)  & -- & -- & -- & -- & $-2.15$ to $0.42$ \\
H$_{\text{GB}}$ (Si-rich) & -- & -- & -- & -- & $0.11$ \\
\bottomrule
\end{tabular}
\vspace{0.5em}
\begin{flushleft}
\small
$^a$Silicon-rich environment. \quad
$^b$Carbon-rich environment. \quad
$^c$LDA calculation; LSDA value in parentheses. \\
$^d$0.90\,eV when the H is centered in the $V_\text{C}$; nudged along the $\langle001\rangle$
  the H migrates to a Si--H--Si at $-0.06$\,eV. \\
$^e$Imposing a $V_\text{C}$ adjacent to the interstitial lowers the formation energy from
  2.93\,eV to 0.02\,eV.
\end{flushleft}
\end{sidewaystable}

The C-rich amorphous structure has the most favorable binding energy at $-2.15$\,eV of all
structures investigated. Four positions were examined, and the binding energies ranged from
$-2.15$ to $0.42$\,eV depending on the local bonding environment. H formed either a C--H bond
or a center-bond (CB) between two Si atoms (Si--H--Si) at an angle of $134.84^\circ$ with bond
lengths of 1.59\,\AA\ and 1.67\,\AA. The CB position was the most favorable with a binding
energy of $-2.33$\,eV, while the C--H bonds resulted in binding energies greater than
$-1.25$\,eV. The lowest C--H binding energy resulted in the shortest bond length of 1.097\,\AA.
The Si-rich amorphous structure resulted in binding energies from $-0.10$ to $0.56$\,eV,
indicating a less energetically favorable position for H. For the Si-rich region, six of the
eight geometry optimizations resulted in a Si--H bond, while the remaining two resulted in a
C--H bond. No general correlation was found between the bond type and the binding energy. See
Table~\ref{tab:solubility} for binding energy results, bond types, and bond lengths.

\subsection{Thermodynamic Stability of H at Finite Temperatures}
\label{sec:thermo_stability}

\begin{figure}[htbp]
  \centering
  \includegraphics[width=\linewidth]{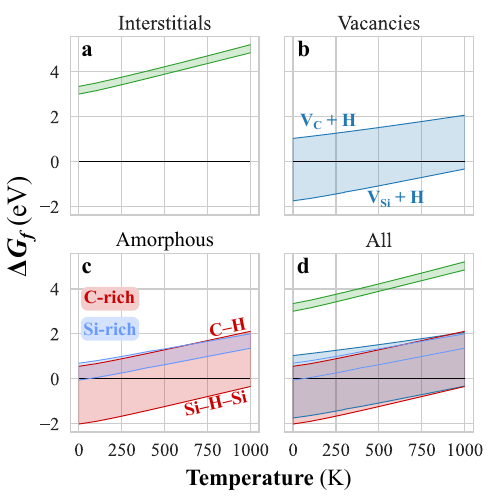}
  \caption{\textit{Ab initio} thermodynamics $\Delta G_f$ results at 1\,bar (0.1\,MPa) from
    0 to 1000\,K for (a)~interstitial sites, (b)~H in carbon and silicon vacancies,
    (c)~carbon-rich (red) and silicon-rich (cyan) amorphous structures, and (d)~superposition of
    all results. (a)~is in reference to the crystalline $\beta$-SiC state. (b,c)~are in
    reference to the defect-induced state. Regions are shaded to visually represent differences
    in free energies.}
  \label{fig:dGf}
\end{figure}

The \textit{ab initio} thermodynamic results indicating the significant impact of temperature and
pressure on the H binding energies are shown in Figure~\ref{fig:dGf}. H in a C-rich amorphous
structure was the most thermodynamically stable, i.e.\ the lowest $\Delta G_f$, for all H
locations investigated, followed by $V_{\text{Si}} + \text{H}$, the Si-rich amorphous structure,
and interstitial H. Interstitial H and $V_{\text{C}} + \text{H}$ both had positive $\Delta G_f$
at all temperatures $\leq1000$\,K, indicating thermodynamic instability and no H trapping in
these configurations. Thus, if both $V_{\text{C}}$ and $V_{\text{Si}}$ are present, H is trapped
in the $V_{\text{Si}}$ at all temperatures, which is in agreement with 0\,K calculations in
Aradi et al.\ \cite{Aradi2001} and Bockstedte et al.\ \cite{Bockstedte2004}.

Amorphous structures presented a wide range of binding sites due to the highly variable
environment. Within the C-rich amorphous structure, H bound to two Si atoms (Si--H--Si) was the
most thermodynamically stable configuration with the lowest $\Delta G_f$ at all temperatures
$\leq1000$\,K, while a C--H bond in the C-rich amorphous structure resulted in higher $\Delta G_f$
values. H within the Si-rich amorphous structure resulted in two bond configurations, a Si--H
bond and a C--H bond. $\Delta G_f$ for both configurations was similar to a C--H bond in a
C-rich amorphous structure; however, all simulations resulted in a positive $\Delta G_f$ except
for temperatures $<100$\,K, indicating that H is not trapped in these positions.

\subsection{Solubility}
\label{sec:solubility_results}

Solubility calculations found that C-rich amorphous structures, Si-rich amorphous structures,
and $V_{\text{Si}}$ are the major contributors to H solubility (including trapping). These
results followed a similar trend to $\Delta G_f$ due to the strong dependence of the solubility
on $\Delta G_f$ (see Eqs.~(S1) and (S2) in SI). Minor contributors include $V_{\text{C}} +
\text{H}$ and interstitial sites adjacent to vacancies, while interstitial sites have effectively
zero solubility. Solubility results, as well as binding energies and free energies of formation,
are summarized in Table~\ref{tab:solubility}. Of all local environments considered, H solubility
is the highest in the C-rich grain boundary.

\begin{table}[htbp]
\centering
\caption{Zero-point corrected 0\,K binding energies, free energies of formation $\Delta G_f$ (at
  600\,K), solubilities (at 600\,K), bond type, and bond length. ``C-rich'' and ``Si-rich'' refer
  to carbon-rich and silicon-rich amorphous region environments, respectively.}
\label{tab:solubility}
\small
\begin{tabular}{lccccc}
\toprule
Defect &
\begin{tabular}[c]{@{}c@{}}Binding\\Energy (eV)\end{tabular} &
\begin{tabular}[c]{@{}c@{}}$\Delta G_f$ (eV)\\at 600\,K\end{tabular} &
\begin{tabular}[c]{@{}c@{}}Solubility\\(mol\,H\,m$^{-3}$)\end{tabular} &
\begin{tabular}[c]{@{}c@{}}Bond\\Type\end{tabular} &
\begin{tabular}[c]{@{}c@{}}Bond\\Length (\AA)\end{tabular} \\
\midrule
T$_{\text{Si}}$ ($4c$)    & 2.87  & 4.08   & $1.24\times10^{-36}$ & None    & -- \\
T$_{\text{C}}$ ($4b$)     & 3.22  & 4.42   & $5.52\times10^{-39}$ & Si--H   & 1.638 \\
(C--H--Si) ($8c$)         & 3.14  & 4.30   & $1.51\times10^{-38}$ & C--H--Si & C--H: 1.088; Si--H: 1.658 \\
(Si--H--Si) ($24f$)       & 2.93  & 4.13   & $8.32\times10^{-38}$ & Si--H--Si & 1.724, 1.722; $179.64^\circ$ \\
$V_{\text{Si}} + \text{H}$   & $-0.97$ & $-0.001$ & $1.64\times10^{-2}$ & C--H  & -- \\
$V_{\text{C}} + \text{H}$    & $-0.06$ & 0.83   & $5.64\times10^{-10}$ & Si--H--Si & 1.615, 1.637; $163.81^\circ$ \\
C-rich (a)  & 0.42   & 1.45  & --                   & C--H      & 1.100 \\
C-rich (b)  & $-0.99$  & 0.07  & $2.73\times10^{-4}$  & C--H      & 1.120 \\
C-rich (c)  & $-1.23$  & $-0.19$ & $2.15\times10^{-2}$  & C--H    & 1.097 \\
C-rich (d)  & $-2.15$  & $-1.06$ & $7.73\times10^{4}$   & Si--H--Si & 1.594, 1.667; $134.84^\circ$ \\
Si-rich (a) & $-0.10$  & 0.87  & $2.24\times10^{-10}$ & Si--H     & 1.513 \\
Si-rich (b) & $-0.10$  & 0.87  & $3.24\times10^{-10}$ & Si--H     & 1.515 \\
Si-rich (c) & $-0.19$  & 0.77  & $2.62\times10^{-9}$  & Si--H     & 1.533 \\
Si-rich (d) & $-0.17$  & 0.86  & $9.23\times10^{-11}$ & C--H      & 1.108 \\
Si-rich (e) & 0.30   & 1.24  & $1.86\times10^{-13}$ & Si--H     & 1.476 \\
Si-rich (f) & 0.56   & 1.46  & $2.93\times10^{-15}$ & Si--H     & 1.518 \\
Si-rich (g) & 0.56   & 1.46  & $5.24\times10^{-15}$ & Si--H     & 1.600 \\
Si-rich (h) & 0.13   & 1.01  & $2.51\times10^{-12}$ & C--H      & 1.091 \\
\bottomrule
\end{tabular}
\end{table}

\begin{figure}[htbp]
  \centering
  \includegraphics[width=\linewidth]{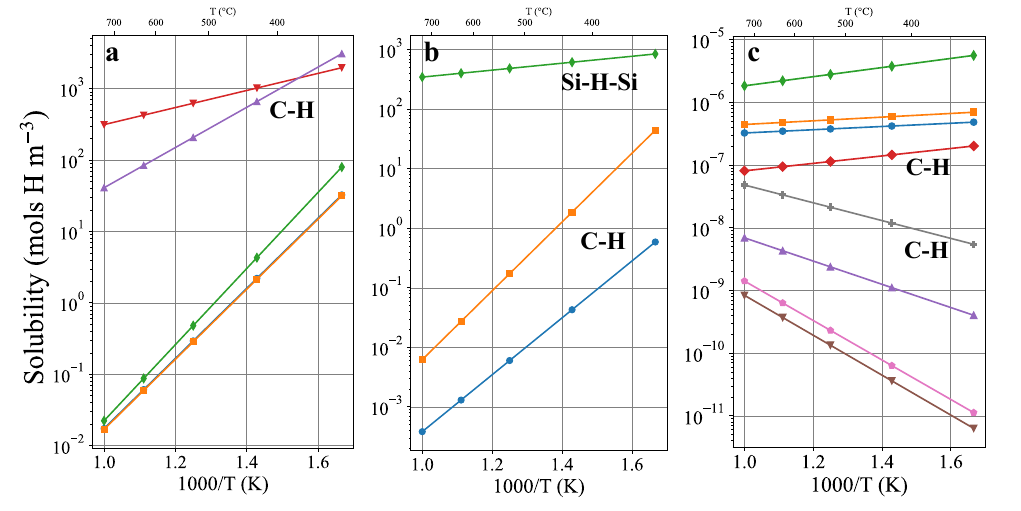}
  \caption{Solubility results for (a)~$V_{\text{Si}}$, (b)~C-rich amorphous structure, and
    (c)~Si-rich amorphous structures. Line labels (e.g., C--H and Si--H--Si) indicate solubility
    results for that bond within the specified structure. In (c), unlabeled lines are solubility
    results for Si--H bonds. Note that the $y$-axis for each panel has a different range.}
  \label{fig:solubility}
\end{figure}

For the C-rich amorphous structure (Figure~\ref{fig:solubility}b), the highest solubility is
found where the H atom formed a Si--H--Si bond at an angle of $134.84^\circ$ and bond lengths of
1.59\,\AA\ and 1.67\,\AA. Lower solubilities were calculated with C--H bonds.
Figure~\ref{fig:solubility}(a,b) also reveals that the C--H bond is more affected by temperature
changes (reflected in its higher slope) compared to the Si--H--Si bond or nonbonded
configuration. For the Si-rich amorphous structures, H solubility was lower than the C-rich
structures and the $V_{\text{Si}}$ (Figure~\ref{fig:solubility}c). Notably, among the eight
calculations, four show an increase in solubility with rising temperature and four show a
decrease. In general, solubility decreases with increasing temperature because the rising thermal
energy enhances the $T\Delta S$ entropy contribution to the free energy, shifting the equilibrium
toward the gas phase rather than dissolution in the solid phase; however, strong binding can
override this for specific configurations. Furthermore, the results demonstrate that
configurations with negative (favorable) binding energies tend to have stable or decreasing
solubility with temperature, whereas those with positive binding energies exhibit increasing
solubility. Similar trends are observed even when comparing lower to higher solubility systems;
for instance, $V_{\text{Si}} + \text{H}$ yields solubilities comparable to those in C-rich
amorphous structures, and a C--H bond in the $V_{\text{Si}} + \text{H}$ complex results in
higher solubility than when H is not bonded.

\section{Discussion}
\label{sec:discussion}

In this work, H in the T$_{\text{Si}}$ had the lowest calculated binding energy of 2.87\,eV
compared to all other interstitial positions, which is in contrast to findings by Aradi et al.\
\cite{Aradi2001}; however, it is in general agreement with a molecular cluster model calculation
found in Ref.~\cite{Roberson1991} and DFT calculations by Kaukonen et al.\ \cite{Kaukonen2003}.
Aradi et al.\ \cite{Aradi2001} found the H equilibrium position to be 0.69\,\AA\ from T$_{\text{C}}$
along the $[111]$ direction, and a metastable configuration 0.08\,\AA\ away from the T$_{\text{Si}}$
site along the $[111]$ direction, while Roberson et al.\ \cite{Roberson1991} molecular cluster
models found H is most stable at T$_{\text{Si}}$. Deak et al.\ \cite{Deak2001} found interstitial H
had the lowest formation energy as a positively charged H atom in an antibonding position behind a
carbon atom. Differences in results are likely due to the different computational methods used.
For example, Aradi et al.\ used LDA with band correction methods, which is known to overestimate
binding energy, compared to this work utilizing the more accurate GGA-PBE functional. In these
calculations, the next lowest interstitial binding energy is 2.93\,eV, where H is located between
two Si atoms (Si--H--Si).

Previous works have shown defects, such as $V_{\text{C}}$ and $V_{\text{Si}}$, and planar
defects, such as grain boundaries, play an important role in permeation
\cite{Verghese1979, Causey1993, Wright2015, Yano1988, Zywietz1999, Zhang2000}. Binding energies
of H in a $V_{\text{Si}}$ are difficult to predict due to multiple local minimum energy
configurations that may exist \cite{Aradi2001}. To capture these effects on solubility, H atoms
were placed in several locations within the $V_{\text{Si}}$. The $V_{\text{Si}}$ is 0.27\,eV
lower than that calculated by Aradi et al.\ \cite{Aradi2001}; however, given the data in
Table~\ref{tab:defects}, it is noted that the $V_{\text{Si}}$ in this work is within one standard
deviation of other works.

H binding in defects is primarily driven by the local electronic environment. In a $V_{\text{C}}$,
the $p$-electrons of the silicon atom provide sufficient extension to bond across the $V_{\text{C}}$,
with a distance of 3.09\,\AA\ that is larger than the 2.35\,\AA\ bond length in pure silicon
\cite{Zywietz1999}. This configuration results in a symmetry-lowering distortion subjecting the
$V_{\text{C}}$ to significant Jahn-Teller distortion \cite{Aradi2001, Zywietz1999}. An analogous
local environment emerges when H is inserted into pure silicon (diamond lattice), where the
distortion is typically a result of H saturating one Si dangling bond, with the remaining two
dangling bonds forming a long bond, pairing the adjacent Si atoms. However, in $\beta$-SiC the H
is inserted symmetrically between two Si atoms forming a Si--H--Si bonding configuration
\cite{Aradi2001, Gali2000}; a configuration that is 0.5\,eV more stable than saturating a single
Si bond. It has been postulated that this is a result of the Si--Si bond distance of 3.1\,\AA\
and the Si--H bond length of 1.5\,\AA, and thus the H atom can interact symmetrically with each
Si atom. The calculations herein support this hypothesis, finding that the remaining Si atoms are
separated by 2.734\,\AA, which is less than the Si--Si distance of 3.085\,\AA\ in $\beta$-SiC,
and the H forms a Si--H--Si with bond lengths of 1.72\,\AA. In contrast, in a $V_{\text{Si}}$
the reduced number of $p$-electrons in the carbon atoms adjacent to the $V_{\text{Si}}$ create
highly localized dangling bonds around the carbon atom. In this local environment, it is expected
that H placed in a $V_{\text{Si}}$ would form a C--H bond; however, in several simulations H
resided in a potential well at a distance of $\approx1.50$\,\AA\ from the C atom, sufficiently
larger than the calculated benchmark C--H bond lengths of 1.082 to 1.185\,\AA, which is
indicative of a nonbonding position \cite{Johnson1999}.

Amorphous regions, which are produced under radiation \cite{Katoh2006, Price1973, Senor2003,
Sprouster2021, Aradi2004, Blackstone1971} and during low-temperature CVD fabrication processes
\cite{Dutta1982, Choi1996}, exhibit a wide range of $\Delta G_f$ and solubilities. These results
are expected as the C-rich structure is effectively a conglomeration of $V_{\text{Si}}$, which
are favored over the $V_{\text{C}}$. It should also be noted that for the C-rich amorphous
structure and $V_{\text{Si}} + \text{H}$ calculations there exists an upper bound at lower
temperatures. This limit is due to all available sites being occupied within the lattice, where
$\theta = 1.0$ (atoms/site) and therefore H interactions become non-negligible. Model accuracies
improve at higher temperatures where a dilute solution is more likely to hold true.

\begin{figure}[htbp]
  \centering
  \includegraphics[width=\linewidth]{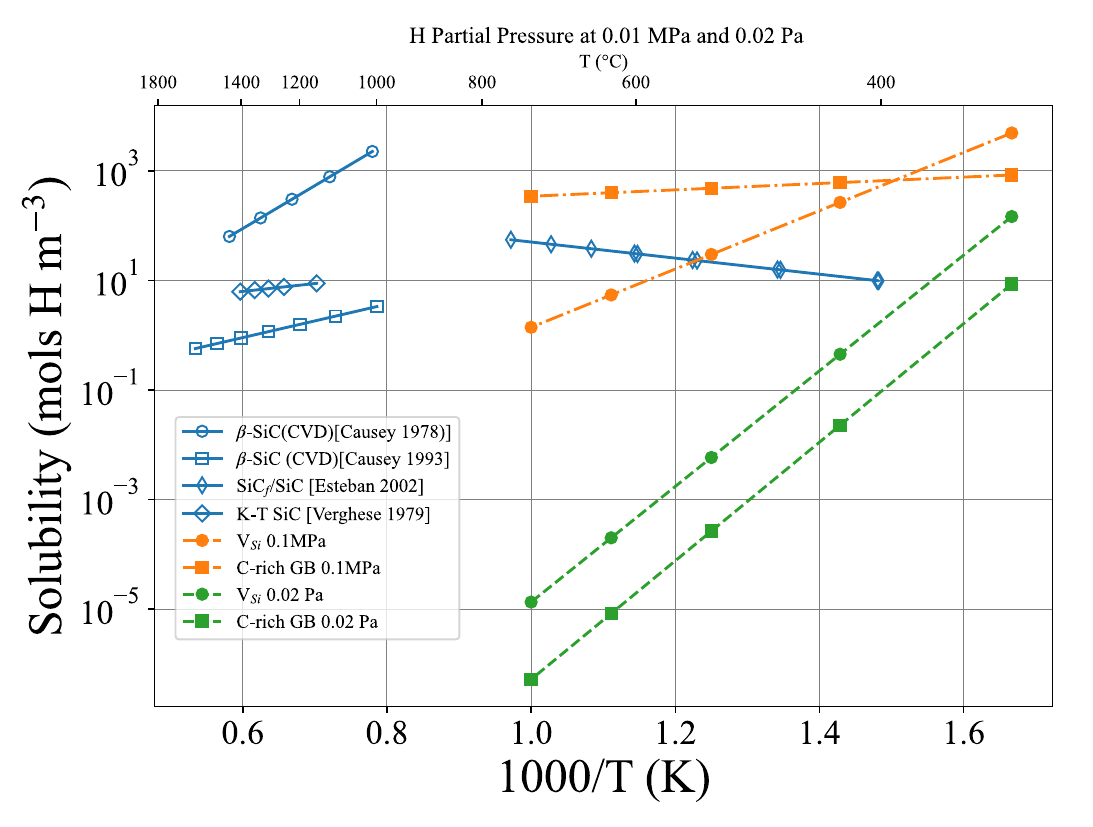}
  \caption{Solubility plot for literature data (open symbols) and calculated solubilities (closed
    symbols) for H in a C-rich amorphous structure and H in a silicon vacancy ($V_{\text{Si}} +
    \text{H}$). Orange and green data represent solubility calculations at 0.1\,MPa (1\,bar) and
    0.02\,Pa ($1\times10^{-4}$\,mbar) H partial pressure.}
  \label{fig:litcomparison}
\end{figure}

These results also have a direct impact on SiC fabrication methods, such as CVD processing,
where variations in processing parameters produce nonstoichiometric regions, which can affect
defect populations. C-rich regions and silicon vacancies, both identified in this work as
dominant contributors to hydrogen solubility, may arise from stoichiometric imbalance during
growth or subsequent irradiation damage. Therefore, controlling deposition parameters to minimize
$V_{\text{Si}}$ concentrations and nonstoichiometric regions represents a promising strategy for
improving tritium permeation barrier performance.

\section{Conclusion}
\label{sec:conclusion}

These results provide quantitative thermodynamic insight into hydrogen behavior in $\beta$-SiC
under conditions relevant to fusion reactor operation. By identifying $V_{\text{Si}}$ and C-rich
amorphous regions as dominant contributors to hydrogen solubility, this work provides a basis for
engineering SiC microstructures with improved tritium permeation resistance and generates
essential inputs for multiscale tritium transport models that couple solubility, diffusivity, and
microstructural evolution.

The findings here demonstrate that H solubility in $\beta$-SiC is highly sensitive to the defect
landscape, local chemical environment, temperature, and H$_2$ partial pressure. Specifically, the
results indicate that stable $V_{\text{Si}}$ in $\beta$-SiC, unlike their metastable counterparts
in 4H-SiC, can lead to increased H solubility. Similar to the $V_{\text{Si}}$, C-rich amorphous
regions significantly increase H solubility. This suggests that tailoring the microstructure—by
minimizing $V_{\text{Si}}$ concentrations and nonstoichiometric C-rich regions—is predicted to
reduce hydrogen solubility and improve permeation barrier performance. However, because our model
currently assumes fixed defect concentrations and neglects contributions from grain boundary
effects, future work should focus on integrating experimental measurements (such as positron
annihilation spectroscopy), exploring variable defect densities, and comparing other statistical
methods to account for the random nature of amorphous regions.

Together, these efforts would provide a more comprehensive understanding of how solubility and
diffusivity combine to control permeation in $\beta$-SiC, ultimately guiding the optimization of
silicon-carbon stoichiometry for developing advanced tritium permeation barriers within the
fusion community.

\section*{Author Contributions (CRediT)}

\noindent
\textbf{Jonathan Evarts:} Writing -- Original Draft, Conceptualization, Formal Analysis,
Methodology, Investigation, Visualization.

\noindent
\textbf{Anne Chaka:} Writing -- Review \& Editing, Conceptualization, Methodology, Supervision.

\noindent
\textbf{Towfiq Ahmed:} Writing -- Review \& Editing, Supervision.

\section*{Declaration of Competing Interest}

The authors declare that they have no known competing financial interests or personal relationships
that could have appeared to influence the work reported in this paper.

\section*{Acknowledgements}

This research was supported by the National Nuclear Security Administration of the US Department
of Energy (DOE) through the Tritium Technology Program at Pacific Northwest National Laboratory.
The authors would like to thank Bruce Schmitt for his insightful and generous conversations
throughout this project, as well as support from Robert Gates.

\section*{Supporting Information}

\textit{Ab initio} Gibbs free energy data and results, modified solubility equations, conversion
factors.

\bibliographystyle{unsrt}
\bibliography{references}

@article{Hassan2021,
  author  = {Hassan, I. A. and Ramadan, H. S. and Saleh, M. A. and Hissel, D.},
  title   = {Hydrogen storage technologies for stationary and mobile applications:
             Review, analysis and perspectives},
  journal = {Renewable Sustainable Energy Rev.},
  volume  = {149},
  pages   = {111311},
  year    = {2021}
}

@article{Ivanov2024,
  author  = {Ivanov, B. V. and Ananev, S. S.},
  title   = {Demand of fusion energy for tritium and the possibility of its production
             in nuclear reactors},
  journal = {At. Energy},
  volume  = {135},
  number  = {3},
  pages   = {183--188},
  year    = {2024}
}

@article{Luscher2013,
  author  = {Luscher, W. G. and Senor, D. J. and Clayton, K. K. and Longhurst, G. R.},
  title   = {In situ measurement of tritium permeation through stainless steel},
  journal = {J. Nucl. Mater.},
  volume  = {437},
  number  = {1},
  pages   = {373--379},
  year    = {2013}
}

@article{Houben2020,
  author  = {Houben, A. and Rasinski, M. and Linsmeier, C.},
  title   = {Hydrogen permeation in fusion materials and the development of tritium
             permeation barriers},
  journal = {Plasma Fusion Res.},
  volume  = {15},
  pages   = {2405016},
  year    = {2020}
}

@article{Linsmeier2017,
  author  = {Linsmeier, C. and Rieth, M. and Aktaa, J. and Chikada, T. and
             Hoffmann, A. and Hoffmann, J. and Houben, A. and Kurishita, H. and
             Jin, X. and Li, M. and Litnovsky, A. and Matsuo, S. and
             {von M\"{u}ller}, A. and Nikolic, V. and Palacios, T. and Pippan, R. and
             Qu, D. and Reiser, J. and Riesch, J. and Shikama, T. and Stieglitz, R. and
             Weber, T. and Wurster, S. and You, J. H. and Zhou, Z.},
  title   = {Development of advanced high heat flux and plasma-facing materials},
  journal = {Nucl. Fusion},
  volume  = {57},
  number  = {9},
  pages   = {092007},
  year    = {2017}
}

@article{Chikada2011,
  author  = {Chikada, T. and Suzuki, A. and Terai, T.},
  title   = {Deuterium permeation and thermal behaviors of amorphous silicon carbide
             coatings on steels},
  journal = {Fusion Eng. Des.},
  volume  = {86},
  number  = {9},
  pages   = {2192--2195},
  year    = {2011}
}

@article{Oya2006,
  author  = {Oya, Y. and Onishi, Y. and Takeda, T. and Kimura, H. and
             Okuno, K. and Tanaka, S.},
  title   = {Interaction between hydrogen isotopes and damaged structures produced by
             {He$^+$} implantation in {SiC}},
  journal = {Fusion Eng. Des.},
  volume  = {81},
  number  = {8},
  pages   = {987--992},
  year    = {2006}
}

@article{Sizyuk2024,
  author  = {Sizyuk, T. and Brooks, J. N. and Abrams, T. and Hassanein, A.},
  title   = {Comprehensive new insights on the potential use of {SiC} as plasma-facing
             materials in future fusion reactors},
  journal = {Nucl. Fusion},
  volume  = {64},
  number  = {8},
  pages   = {086036},
  year    = {2024}
}

@article{Causey1978,
  author  = {Causey, R. A. and Fowler, J. D. and Ravanbakht, C. and
             Elleman, T. S. and Verghese, K.},
  title   = {Hydrogen diffusion and solubility in silicon carbide},
  journal = {J. Am. Ceram. Soc.},
  volume  = {61},
  number  = {5--6},
  pages   = {221--225},
  year    = {1978}
}

@article{Causey1995,
  author  = {Causey, R. A. and Wampler, W. R.},
  title   = {The use of silicon carbide as a tritium permeation barrier},
  journal = {J. Nucl. Mater.},
  volume  = {220--222},
  pages   = {823--826},
  year    = {1995}
}

@article{Verghese1979,
  author  = {Verghese, K. and Zumwalt, L. R. and Feng, C. P. and Elleman, T. S.},
  title   = {Hydrogen permeation through non-metallic solids},
  journal = {J. Nucl. Mater.},
  volume  = {85--86},
  pages   = {1161--1164},
  year    = {1979}
}

@article{Sinharoy1984,
  author  = {Sinharoy, S. and Lange, W. J.},
  title   = {Summary abstract: Permeation of hydrogen through {CVD} silicon carbide},
  journal = {J. Vac. Sci. Technol. A},
  volume  = {2},
  number  = {2},
  pages   = {636--637},
  year    = {1984}
}

@article{Causey1993,
  author  = {Causey, R. A. and Wampler, W. R. and Retelle, J. R. and Kaae, J. L.},
  title   = {Tritium migration in vapor-deposited $\beta$-silicon carbide},
  journal = {J. Nucl. Mater.},
  volume  = {203},
  number  = {3},
  pages   = {196--205},
  year    = {1993}
}

@incollection{Causey2012,
  author    = {Causey, R. A. and Karnesky, R. A. and {San Marchi}, C.},
  title     = {Tritium barriers and tritium diffusion in fusion reactors},
  booktitle = {Comprehensive Nuclear Materials},
  editor    = {Konings, R. J. M.},
  publisher = {Elsevier},
  address   = {Oxford},
  pages     = {511--549},
  year      = {2012}
}

@article{Tam1995,
  author  = {Tam, S. W. and Kopasz, J. P. and Johnson, C. E.},
  title   = {Tritium transport and retention in {SiC}},
  journal = {J. Nucl. Mater.},
  volume  = {219},
  pages   = {87--92},
  year    = {1995}
}

@inproceedings{Minami2007,
  author    = {Minami, T. and Niigawa, S. and Ueno, Y. and Hinoki, T. and
               Yamamoto, Y. and Konishi, S.},
  title     = {{SiC} permeation study},
  booktitle = {Proc.\ 2007 IEEE 22nd Symp.\ on Fusion Engineering},
  pages     = {1--4},
  year      = {2007}
}

@article{Yamamoto2016,
  author  = {Yamamoto, Y. and Murakami, Y. and Yamaguchi, H. and Yamamoto, T. and
             Yonetsu, D. and Noborio, K. and Konishi, S.},
  title   = {Re-evaluation of {SiC} permeation coefficients at high temperatures},
  journal = {Fusion Eng. Des.},
  volume  = {109--111},
  pages   = {1286--1290},
  year    = {2016}
}

@article{Wright2015,
  author  = {Wright, G. M. and Durrett, M. G. and Hoover, K. W. and
             Kesler, L. A. and Whyte, D. G.},
  title   = {Silicon carbide as a tritium permeation barrier in tungsten
             plasma-facing components},
  journal = {J. Nucl. Mater.},
  volume  = {458},
  pages   = {272--274},
  year    = {2015}
}

@article{Katoh2006,
  author  = {Katoh, Y. and Hashimoto, N. and Kondo, S. and Snead, L. L. and Kohyama, A.},
  title   = {Microstructural development in cubic silicon carbide during irradiation at
             elevated temperatures},
  journal = {J. Nucl. Mater.},
  volume  = {351},
  number  = {1},
  pages   = {228--240},
  year    = {2006}
}

@article{Price1973,
  author  = {Price, R. J.},
  title   = {Neutron irradiation-induced voids in $\beta$-silicon carbide},
  journal = {J. Nucl. Mater.},
  volume  = {48},
  number  = {1},
  pages   = {47--57},
  year    = {1973}
}

@article{Senor2003,
  author  = {Senor, D. J. and Youngblood, G. E. and Greenwood, L. R. and
             Archer, D. V. and Alexander, D. L. and Chen, M. C. and Newsome, G. A.},
  title   = {Defect structure and evolution in silicon carbide irradiated to
             1 dpa-{SiC} at 1100\,{\textdegree C}},
  journal = {J. Nucl. Mater.},
  volume  = {317},
  number  = {2},
  pages   = {145--159},
  year    = {2003}
}

@article{Sprouster2021,
  author  = {Sprouster, D. J. and Koyanagi, T. and Drey, D. L. and
             Katoh, Y. and Snead, L. L.},
  title   = {Atomic and microstructural origins of stored energy release in
             neutron-irradiated silicon carbide},
  journal = {Phys. Rev. Mater.},
  volume  = {5},
  number  = {10},
  pages   = {103601},
  year    = {2021}
}

@incollection{Snead2012,
  author    = {Snead, L. L. and Katoh, Y. and Nozawa, T.},
  title     = {Radiation effects in {SiC} and {SiC--SiC}},
  booktitle = {Comprehensive Nuclear Materials},
  editor    = {Konings, R. J. M.},
  publisher = {Elsevier},
  address   = {Oxford},
  pages     = {215--240},
  year      = {2012}
}

@article{Sun2017,
  author  = {Sun, J. and You, Y.-W. and Hou, J. and Li, X. and Li, B. S. and
             Liu, C. S. and Wang, Z. G.},
  title   = {The effect of irradiation-induced point defects on energetics and kinetics
             of hydrogen in 3C-{SiC} in a fusion environment},
  journal = {Nucl. Fusion},
  volume  = {57},
  number  = {6},
  pages   = {066031},
  year    = {2017}
}

@article{Yano1988,
  author  = {Yano, T. and Suzuki, T. and Maruyama, T. and Iseki, T.},
  title   = {Microstructure and annealing behavior of heavily neutron-irradiated
             $\beta$-{SiC}},
  journal = {J. Nucl. Mater.},
  volume  = {155--157},
  pages   = {311--314},
  year    = {1988}
}

@article{Esteban2002,
  author  = {Esteban, G. A. and Perujo, A. and Legarda, F. and
             Sedano, L. A. and Riccardi, B.},
  title   = {Deuterium transport in {SiC}$_\mathrm{f}$/{SiC} composites},
  journal = {J. Nucl. Mater.},
  volume  = {307--311},
  pages   = {1430--1435},
  year    = {2002}
}

@article{Aradi2001,
  author  = {Aradi, B. and Gali, A. and De\'{a}k, P. and Lowther, J. E. and
             Son, N. T. and Janz\'{e}n, E. and Choyke, W. J.},
  title   = {\textit{Ab initio} density-functional supercell calculations of hydrogen
             defects in cubic {SiC}},
  journal = {Phys. Rev. B},
  volume  = {63},
  number  = {24},
  pages   = {245202},
  year    = {2001}
}

@article{Bernardini2004,
  author  = {Bernardini, F. and Mattoni, A. and Colombo, L.},
  title   = {Energetics of native point defects in cubic silicon carbide},
  journal = {Eur. Phys. J. B},
  volume  = {38},
  number  = {3},
  pages   = {437--444},
  year    = {2004}
}

@incollection{Iwamoto2015,
  author    = {Iwamoto, N. and Svensson, B. G.},
  title     = {Point defects in silicon carbide},
  booktitle = {Semiconductors and Semimetals},
  editor    = {Romano, L. and Privitera, V. and Jagadish, C.},
  volume    = {91},
  publisher = {Elsevier},
  pages     = {369--407},
  year      = {2015}
}

@article{Kaukonen2003,
  author  = {Kaukonen, M. and Fall, C. J. and Lento, J.},
  title   = {Interstitial {H} and {H}$_2$ in {SiC}},
  journal = {Appl. Phys. Lett.},
  volume  = {83},
  number  = {5},
  pages   = {923--925},
  year    = {2003}
}

@article{Torpo2001,
  author  = {Torpo, L. and Marlo, M. and Staab, T. E. M. and Nieminen, R. M.},
  title   = {Comprehensive \textit{ab initio} study of properties of monovacancies and
             antisites in 4H-{SiC}},
  journal = {J. Phys.: Condens. Matter},
  volume  = {13},
  number  = {28},
  pages   = {6203},
  year    = {2001}
}

@article{Wang2020,
  author  = {Wang, X. and Zhao, J. and Xu, Z. and Djurabekova, F. and
             Rommel, M. and Song, Y. and Fang, F.},
  title   = {Density functional theory calculation of the properties of carbon vacancy
             defects in silicon carbide},
  journal = {Nanotechnol. Precis. Eng.},
  volume  = {3},
  number  = {4},
  pages   = {211--217},
  year    = {2020}
}

@article{Zywietz1999,
  author  = {Zywietz, A. and Furthm\"{u}ller, J. and Bechstedt, F.},
  title   = {Vacancies in {SiC}: Influence of {Jahn-Teller} distortions, spin effects,
             and crystal structure},
  journal = {Phys. Rev. B},
  volume  = {59},
  number  = {23},
  pages   = {15166--15180},
  year    = {1999}
}

@inproceedings{Nadaoka2009,
  author    = {Nadaoka, R. and Uriu, K. and Yamamoto, Y. and Konishi, S.},
  title     = {{SiC} hydrogen solubility study},
  booktitle = {Proc.\ 2009 23rd IEEE/NPSS Symp.\ on Fusion Engineering},
  pages     = {1--4},
  year      = {2009}
}

@article{Delley2000,
  author  = {Delley, B.},
  title   = {From molecules to solids with the {DMol}$^3$ approach},
  journal = {J. Chem. Phys.},
  volume  = {113},
  number  = {18},
  pages   = {7756--7764},
  year    = {2000}
}

@article{Brauer1996,
  author  = {Brauer, G. and Anwand, W. and Coleman, P. and Knights, A. and
             Plazaola, F. and Pacaud, Y. and Skorupa, W. and St\"{o}rmer, J. and
             Willutzki, P.},
  title   = {Positron studies of defects in ion-implanted {SiC}},
  journal = {Phys. Rev. B},
  volume  = {54},
  number  = {5},
  pages   = {3084},
  year    = {1996}
}

@article{Lee2015,
  author  = {Lee, K. and Yuan, M. and Wilcox, J.},
  title   = {Understanding deviations in hydrogen solubility predictions in transition
             metals through first-principles calculations},
  journal = {J. Phys. Chem. C},
  volume  = {119},
  number  = {34},
  pages   = {19642--19653},
  year    = {2015}
}

@book{Chase1998,
  author    = {Chase, Malcolm W., Jr.},
  title     = {{NIST-JANAF} Thermochemical Tables},
  edition   = {4th},
  publisher = {American Chemical Society / American Institute of Physics},
  address   = {Washington, DC / New York},
  year      = {1998}
}

@article{Chaka2014,
  author  = {Chaka, A. M. and Felmy, A. R.},
  title   = {\textit{Ab Initio} thermodynamic model for magnesium carbonates and hydrates},
  journal = {J. Phys. Chem. A},
  volume  = {118},
  number  = {35},
  pages   = {7469--7488},
  year    = {2014}
}

@article{Reuter2001,
  author  = {Reuter, K. and Scheffler, M.},
  title   = {Composition, structure, and stability of {RuO}$_2$(110) as a function of
             oxygen pressure},
  journal = {Phys. Rev. B},
  volume  = {65},
  number  = {3},
  pages   = {035406},
  year    = {2001}
}

@article{Wang2000,
  author  = {Wang, X. and Chaka, A. and Scheffler, M.},
  title   = {Effect of the environment on $\alpha$-{Al}$_2${O}$_3$(0001) surface structures},
  journal = {Phys. Rev. Lett.},
  volume  = {84},
  number  = {16},
  pages   = {3650--3653},
  year    = {2000}
}

@article{Zhang2004,
  author  = {Zhang, W. and Smith, J. R. and Wang, X. G.},
  title   = {Thermodynamics from \textit{ab initio} computations},
  journal = {Phys. Rev. B},
  volume  = {70},
  number  = {2},
  pages   = {024103},
  year    = {2004}
}

@article{Mason2010,
  author  = {Mason, S. E. and Iceman, C. R. and Trainor, T. P. and Chaka, A. M.},
  title   = {Density functional theory study of clean, hydrated, and defective alumina
             (1$\bar{1}$02) surfaces},
  journal = {Phys. Rev. B},
  volume  = {81},
  number  = {12},
  pages   = {125423},
  year    = {2010}
}

@article{Bockstedte2004,
  author  = {Bockstedte, M. and Mattausch, A. and Pankratov, O.},
  title   = {\textit{Ab initio} study of the annealing of vacancies and interstitials in
             cubic {SiC}: Vacancy-interstitial recombination and aggregation of carbon
             interstitials},
  journal = {Phys. Rev. B},
  volume  = {69},
  number  = {23},
  pages   = {235202},
  year    = {2004}
}

@article{Roberson1991,
  author  = {Roberson, M. A. and Estreicher, S. K.},
  title   = {Interstitial hydrogen in cubic and hexagonal {SiC}},
  journal = {Phys. Rev. B},
  volume  = {44},
  number  = {19},
  pages   = {10578--10584},
  year    = {1991}
}

@article{Deak2001,
  author  = {De\'{a}k, P. and B\'{a}lint, A. and Adam, G.},
  title   = {Boron and aluminium doping in {SiC} and its passivation by hydrogen},
  journal = {J. Phys.: Condens. Matter},
  volume  = {13},
  number  = {40},
  pages   = {9019},
  year    = {2001}
}

@article{Zhang2000,
  author  = {Zhang, X. F. and Sixta, M. E. and {De Jonghe}, L. C.},
  title   = {Grain boundary evolution in hot-pressed {ABC-SiC}},
  journal = {J. Am. Ceram. Soc.},
  volume  = {83},
  number  = {11},
  pages   = {2813--2820},
  year    = {2000}
}

@article{Gali2000,
  author  = {Gali, A. and Aradi, B. and De\'{a}k, P. and Choyke, W. J. and Son, N. T.},
  title   = {Overcoordinated hydrogens in the carbon vacancy: Donor centers of {SiC}},
  journal = {Phys. Rev. Lett.},
  volume  = {84},
  number  = {21},
  pages   = {4926--4929},
  year    = {2000}
}

@misc{Johnson1999,
  author = {Johnson, R.},
  title  = {{NIST} 101: Computational Chemistry Comparison and Benchmark Database},
  note   = {Personal correspondence},
  year   = {1999}
}

@article{Aradi2004,
  author  = {Aradi, B. and De\'{a}k, P. and Gali, A. and Son, N. T. and Janz\'{e}n, E.},
  title   = {Diffusion of hydrogen in perfect, $p$-type doped, and radiation-damaged
             4H-{SiC}},
  journal = {Phys. Rev. B},
  volume  = {69},
  number  = {23},
  pages   = {233202},
  year    = {2004}
}

@article{Blackstone1971,
  author  = {Blackstone, R. and Voice, E. H.},
  title   = {The expansion of silicon carbide by neutron irradiation at high temperature},
  journal = {J. Nucl. Mater.},
  volume  = {39},
  number  = {3},
  pages   = {319--322},
  year    = {1971}
}

@article{Dutta1982,
  author  = {Dutta, R. and Banerjee, P. K. and Mitra, S. S.},
  title   = {Effect of hydrogenation on the electrical conductivity of amorphous silicon
             carbide},
  journal = {Solid State Commun.},
  volume  = {42},
  number  = {3},
  pages   = {219--222},
  year    = {1982}
}

@article{Choi1996,
  author  = {Choi, W. K. and Loo, F. L. and Loh, F. C. and Tan, K. L.},
  title   = {Effects of hydrogen and rf power on the structural and electrical properties
             of rf sputtered hydrogenated amorphous silicon carbide films},
  journal = {J. Appl. Phys.},
  volume  = {80},
  number  = {3},
  pages   = {1611--1616},
  year    = {1996}
}

\end{document}